\documentclass[a4paper,11pt]{article} 
\usepackage[german,english]{babel} 
\usepackage{a4wide,graphicx,float, hyperref, amsmath,amssymb} 
\usepackage[authoryear,round,longnamesfirst]{natbib}

\newcommand{\Expect}{\mbox{\textnormal{E}}}
\newcommand{\var}{\mbox{\textnormal{Var}}}

\let\proglang=\textsf
\newcommand{\pkg}[1]{{\fontseries{b}\selectfont #1}}

\RequirePackage[T1]{fontenc} 
\RequirePackage{ae,fancyvrb}
\DefineVerbatimEnvironment{Sinput}{Verbatim}{fontshape=sl}
\DefineVerbatimEnvironment{Soutput}{Verbatim}{}
\DefineVerbatimEnvironment{Scode}{Verbatim}{fontshape=sl}

\setkeys{Gin}{width=0.9\textwidth}

\begin{document} 
\renewcommand{\baselinestretch}{1.0} 

\title{\bf Structural Change in (Economic) Time Series}
 
\author{\hfill Christian Kleiber 
\thanks{Date: April 25, 2016. \newline Correspondence: Christian Kleiber, 
Faculty of Business and Economics, Universit\"at Basel, 
Peter Merian-Weg 6, CH-4002 Basel, Switzerland. 
E-mail: \texttt{christian.kleiber@unibas.ch} } 
} 

\date{}
\maketitle

\begin{abstract} 
Methods for detecting structural changes, or change points, in time series data are widely used in 
many fields of science and engineering. This chapter sketches some basic methods for the analysis of 
structural changes in time series data. The exposition is confined to retrospective methods for 
univariate time series. Several recent methods for dating structural changes are compared using a 
time series of oil prices spanning more than 60 years. The methods broadly agree for the first part 
of the series up to the mid-1980s, for which changes are associated with major historical events,   
but provide somewhat different solutions thereafter, reflecting a gradual increase in oil prices 
that is not well described by a step function. As a further illustration, 1990s data on the 
volatility of the Hang Seng stock market index are reanalyzed. 

\smallskip


\smallskip

\end{abstract}

\section{Introduction}

In time series analysis, the point of reference is that of a stationary stochastic process; i.e., a process for which the sequence of first and second-order moments is constant (`weak stationarity'), or even the sequence of the entire marginal distributions (`strict stationarity'). In practice, many time series exhibit some form of nonstationarity: changing levels, changing variances, changing autocorrelations, or a combination of some or all of these aspects. These phenomena are then called structural changes or structural breaks and the associated statistical methodology is sometimes called change point analysis. Such phenomena may be seen as `complex' in the sense of this Volume, in that classical models with constant coefficients are rejected by the data. 

Structural change methodology is widely used in economics, finance, bioinformatics, 
engineering, public health, and climatology, 
to mention just a few fields of application. 
An interesting recent contribution \citep{compsyn:Kelly+OGrada:2014} disputes the existence 
of a `little ice age' for parts of Central and Northern Europe between the 14th and 19th century. 
Using structural change methodology (of the type used below) on 
temperature reconstructions spanning several centuries, 
Kelly and \'{O}~Gr\'{a}da find no evidence for sustained falls in mean temperatures prior to 1900, 
instead several relevant series are best seen as white noise series. 
One explanation for the contradiction to the established view is the 
climatological practice of smoothing data prior to analysis. When the raw data are in fact uncorrelated, such preprocessing can introduce spurious dependencies (the `Slutsky effect').

More than 25 years ago, a bibliography on structural change methodology and applications 
published in the economics literature \citep{compsyn:Hackl+Westlund:1989} 
already lists some 500 references, and the literature has grown rather rapidly since then. More recently, a bibliography available with the 
\textsf{R} package \textbf{strucchange} provides more than 800 references, ending in 2006 \citep{compsyn:Zeileis+Leisch+Hornik:2002}. 
Recent surveys of the methodology 
include \cite{compsyn:Perron:2006}, \cite{compsyn:Aue+Horvath:2013} and \cite{compsyn:Horvath+Rice:2014}. 
Much of this methodology relies quite heavily on functional central limit theorems (FCLTs), an excellent reference is 
\cite{compsyn:Csorgo+Horvath:1997}.

Apart from practical relevance of the associated issues, one reason for the large number of publications is that the notion of `structural change' can be formalized in many different ways. In terms of statistical hypothesis tests, the null hypothesis of `no structural change' is reasonably clear (model parameters are constant), but the alternative can mean many things: a parameter (or several parameters) change(s) its (their) value(s) abruptly 
(once, twice, or more often), or it changes gradually according to a stochastic mechanism (e.g., via a random coefficient model), or it switches randomly among a small number of states (e.g., via a hidden Markov model), etc. There are many further possibilities.

The available methodology therefore incorporates ideas from a variety of fields: linear models, sequential analysis, wavelets, etc. In economics, there is comparatively greater interest in changes in regression models, whereas in many other fields of application interest is focused on changes in a univariate time series. A further dividing line is on-line (sequential) analysis of a growing sample vs. off-line (retrospective) analysis of a fixed sample.

This chapter provides some basic ideas of  
change point methodology along with empirical examples. It is biased towards least-squares methods, methodology used in economics and finance as well as availability in statistical software. 
The following section outlines selected methods in the context of a simple signal-plus-noise model. For reasons of space, the exposition is confined to retrospective analysis of changes in univariate time series. 
In section~\ref{sec:oilprice}, several recent algorithms are explored for dating structural changes in a series of oil prices. Section~\ref{sec:financial} dates volatility changes in Hang Seng stock market index returns, thereby revisiting data formerly studied in \cite{compsyn:Andreou+Ghysels:2002}.  
The final section provides some references for sequential analysis of structural change and 
also for more complex data structures.

\section{Some basic ideas in change point analysis}\label{sec:basic}

To fix ideas, consider a signal-plus-noise model for an observable (univariate) quantity $y_i$, 

\[
y_i ~=~ \mu_i + e_i , \quad i = 1, \dots, T,
\]

where $\mu_i$ is the (deterministic) signal and $e_i$ is the noise, with $\Expect[e_i] = 0$ and $\var[e_i] = \sigma^2$. 
As noted above, this chapter is confined to changes in a univariate time series. However, for many methods, there is a regression version with $\mu_i = x_i ^\top \beta_i$, 
where $x_i$ is a vector of covariates and $\beta_i$ the corresponding set of regression coefficients. 
In the classical setting, the $e_i$ form a sequence of independent and identically distributed (i.i.d.) random variables, 
but many more recent contributions, especially in economics and finance, consider dependent processes.

\subsection{Testing for structural change}

In terms of statistical testing, the null hypothesis of interest is $H_0: \mu_i = \mu_0$ for all $i$; i.e., the signal exhibits no change. (For the regression version, the corresponding null hypothesis is $H_0: \beta_i = \beta_0$ for all $i$.) 
Under the null hypothesis, natural estimates of $\mu_0$ are the recursive estimates $\hat \mu_k = k^{-1} \sum_{i=1}^k y_i $, $k= 1, \dots, T$; i.e., the sequence of sample means computed from a growing sample. 
The corresponding recursive residuals are $\tilde e_i = y_i - \hat \mu_{i-1}$, $i = 2, \dots, T$. 

A classical idea is to study the fluctuations of partial (or cumulative) sums (CUSUMs) of these recursive residuals and to reject the null hypothesis of parameter stability if their fluctuations are excessive. 
In order to assess significance, introduce an \emph{empirical fluctuation process} 
indexed by $t \in [0,1]$ as the process of partial sums of the recursive residuals via

\begin{equation*}
\tilde S_T(t) ~=~ \frac{1}{\hat \sigma_T \sqrt{T}} \sum_{i=1}^{[Tt]} \tilde e_i, \quad 0 \leq t \leq 1 ,
\end{equation*}

where $[Tt]$ denotes the integer part of $Tt$ and $\hat \sigma_T^2$ some consistent estimate 
of $\sigma^2$. This object is often called the 
\emph{Rec-CUSUM process} as it is based on recursive residuals. 
It is well known that under the above assumptions this empirical fluctuation process 
can be approximated by a Brownian motion, $B(t)$, $0 \leq t \leq 1$, 
and hence the enormous literature on properties of this stochastic process 
can be used to assess the fluctuations in the recursive residuals. Excessive fluctuation 
is determined from the crossing probabilities of certain boundaries, this is the approach proposed 
in the seminal paper by \cite{compsyn:Brown+Durbin+Evans:1975}.  

However, from a regression point of view, the ordinary least-squares (OLS) residuals 
$\hat e_i = y_i - \hat \mu_T$  are perhaps a more natural starting point, 
leading to the test statistic

\begin{equation}\label{olscusum}
\max _{k = 1, \dots, T} \left| \frac{1}{\hat \sigma_T \sqrt{T}} \sum_{i=1}^k \hat e_i \right| .
\end{equation}

As OLS residuals are correlated and sum to zero by construction, 
the limiting process corresponding to the \emph{OLS-CUSUM process} 
is no longer a Brownian motion. 
Instead, the limiting process is now a Brownian bridge, $B^0(t)$, 
with $B^0(t) = B(t) - t~B(1)$, $0 \leq t \leq 1$, and the relevant limiting quantity 
for assessing significant deviation from the null hypothesis is

\begin{equation*}
\sup_{0 \leq t \leq 1} |B^0(t)| ,
\end{equation*}

the supremum of the absolute value of a Brownian bridge on the unit interval  
\citep{compsyn:Ploberger+Kraemer:1992}. 
This object is well known in the statistical literature, and quantiles of its 
distribution provide critical values for a test based on (\ref{olscusum}). 

To briefly illustrate the machinery, consider a time series of measurements of the annual flow 
of the river Nile at Aswan, for the period 1871 to 1970. 
This series is part of any binary distribution of 
\proglang{R}~\citep{compsyn:R:2016} under the name \texttt{Nile} 
and has been used repeatedly in the statistical literature on change point methods. 
The following illustrations make use of the \textsf{R} package \textbf{strucchange}, 
which among other things implements 
structural change detection using empirical fluctuation processes and related techniques. 
It should be noted that the original paper \citep{compsyn:Zeileis+Leisch+Hornik:2002} 
describing the software documents the first release of the package, 
but many methods were added in subsequent years, including the methods for dating structural changes that are used in the next section. The package is still actively maintained, but the main developments happened some 10 years ago. 

\begin{figure}[t!]
\includegraphics{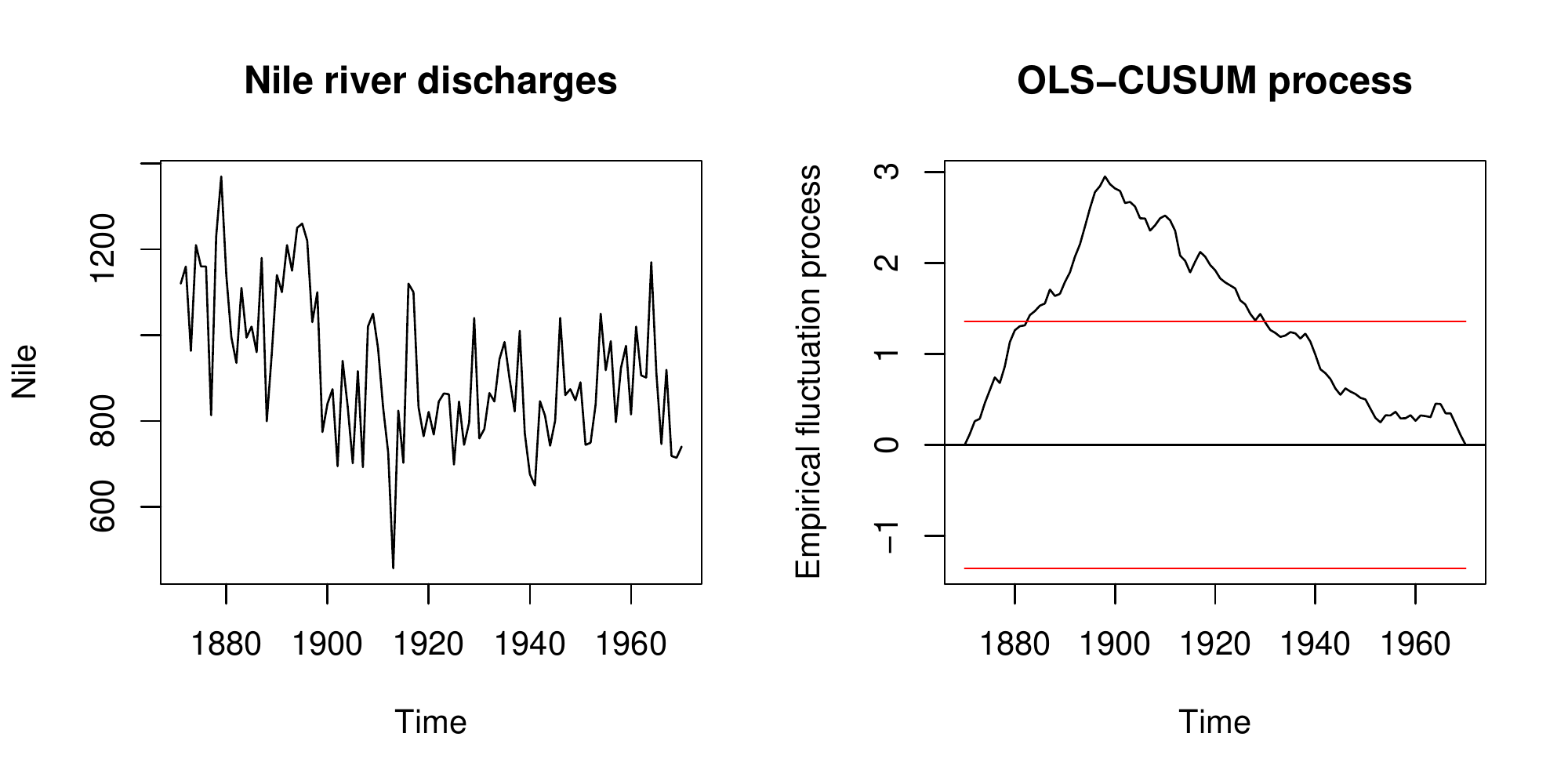}

\caption{\label{fig:niledata} Nile river discharges: Raw data (left) and OLS-CUSUM process (right).}
\end{figure}

Figure~\ref{fig:niledata} plots the time series (left panel) and the corresponding 
OLS-CUSUM process (right panel) along with a boundary indicating the 5\%~critical value 
for the test statistic (\ref{olscusum}). 
It is seen that the empirical fluctuation process crosses the boundary, and hence 
the hypothesis of a constant level is rejected at the 5\%~level.

The test statistic (\ref{olscusum}), often called the OLS-CUSUM test, measures the 
maximal absolute deviation from zero of the corresponding OLS-CUSUM process. 
There are many variations of this idea. For example, it is possible to use other functionals 
of the empirical fluctuation process, such as the range or some average of the fluctuations. 
It is also possible to study moving instead of cumulative sums, leading to moving sum (MOSUM) processes. 
Or, instead of the fluctuations in the residuals, one can directly assess the fluctuations 
in the estimates themselves; there are again recursive and moving versions \citep{compsyn:Kuan+Hornik:1995}. 
In the univariate case considered here, the latter idea is equivalent to CUSUMs or MOSUMs 
of the residuals, but in the regression case it leads to new procedures. 
It is also possible to assess fluctuations in first-order conditions of fitting methods 
other than least squares, for example likelihood methods \citep{compsyn:Zeileis:2005}.

Also, applications in economics and finance often involve dependent data, so that the machinery described above requires adjustments. These involve the \emph{long-run variance},

\[
\omega^2 ~=~ \lim_{T\to\infty} \var(\tilde S_T(1)) .
\]

If a consistent estimator $\hat \omega^2$ of $\omega^2$ is available, 
then $\hat\omega^{-1}\tilde S_T(t) $ or $\hat\omega^{-1}\hat S_T(t) $ can, in many settings of interest, 
again be approximated by a Brownian motion or a Brownian bridge.

\subsection{Dating structural changes}

Having found evidence for the presence of structural change it is of interest to estimate the change points themselves. 
In economics, basic references for dating structural changes are \cite{compsyn:Bai+Perron:1998, compsyn:Bai+Perron:2003}, which among other things provide a method for obtaining confidence intervals for the break dates. The associated point estimation issue -- the segmentation of the sample into homogeneous parts -- dates back at least to  \cite{compsyn:Bellman+Roth:1969}.

The model of interest is now a step function for the signal. With $m+1$ segments (corresponding to $m$ breaks), this is

\begin{equation} \label{datingmodel}
y_i ~=~ \mu_j + e_i , \qquad \tau_{j-1} + 1 ~ \leq ~ i ~ \leq ~ \tau_j, \quad j = 1, \dots, m+1. 
\end{equation} 

Here $j$ is the segment index and $\{\tau_1, \dots, \tau_m\}$ denotes the set of the break points. By convention, $\tau_0 = 0$ and $\tau_{m+1} = T$. 

Given the break points $\tau_1, \dots, \tau_m$, the least squares estimate of $\mu_j$ 
is the sample mean of the observations pertaining to segment~$j$.  
The resulting aggregate residual sum of squares is given by

\begin{equation} \label{RSSrss}
RSS(\tau_1, \dots, \tau_m) \quad = \quad \sum_{j = 1}^{m+1} rss(\tau_{j-1} + 1, \tau_j),
\end{equation}

where $rss(\tau_{j-1} + 1, \tau_j)$ 
is the residual sum of squares for  segment~$j$.
The problem of dating structural changes is to find the break points $\hat{\tau}_1, \dots,
\hat{\tau}_m$ that minimize the objective function, 

\begin{equation} \label{argmin}
\{\hat{\tau}_1, \dots, \hat{\tau}_m\} \quad 
= \quad \mbox{argmin}_{\{\tau_1, \dots, \tau_m\}} RSS(\tau_1, \dots, \tau_m) , 
\end{equation}

over all partitions $\{\tau_1, \dots, \tau_m\}$ with $\tau_j - \tau_{j-1} \ge T_{\min}$. 
Here $T_{\min}$ is a bandwidth parameter to be specified by the user, it defines the minimal segment length. 
For a given number $m$ of break points, their optimal location can be found using a 
dynamic programming algorithm. The number of change points $m$ itself can be determined 
via information criteria, the \texttt{breakpoints()} function in \pkg{strucchange} employs the 
Bayesian Information Criterion (BIC). Below, this algorithm is referred to as \texttt{breakpoints}. 
More details on the implementation may be found in \cite{compsyn:Zeileis+Kleiber+Kraemer:2003}.

\begin{figure}[t!]
\includegraphics{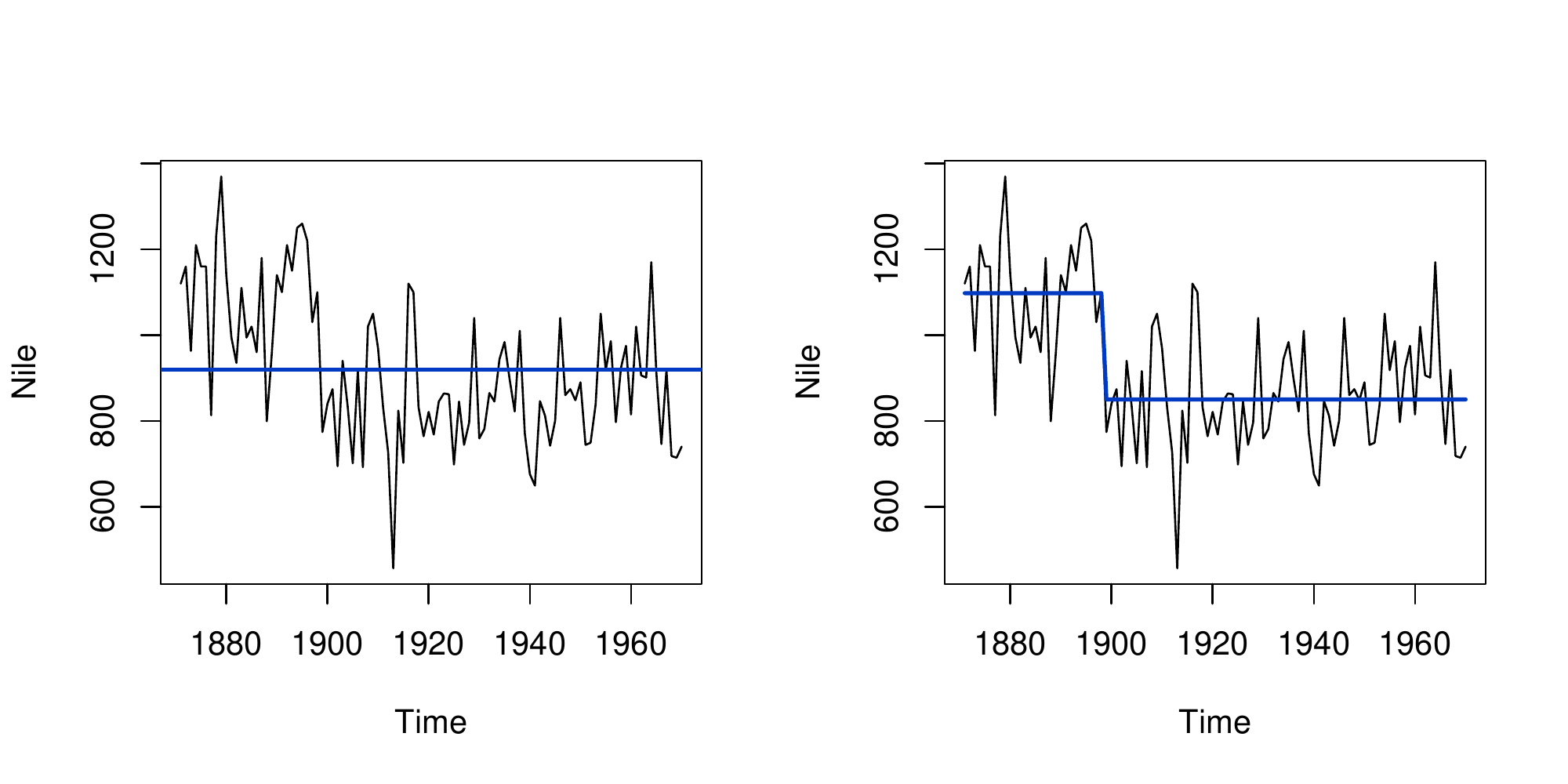}

\caption{\label{fig:nilebreak} Nile river discharges: Model with constant level (left) vs. model with two regimes (right).}
\end{figure}

\medskip

Returning to the Nile river flows, Figure~\ref{fig:nilebreak} (left panel) provides the fit 
for a traditional autoregressive model of order one (AR(1)) with constant parameters. 
Clearly, the fit is quite poor in that for the first part of the series the data are almost always above the fitted mean level, whereas after approximately the year 1900 they are mostly below. 
In contrast, Figure~\ref{fig:nilebreak} (right panel) provides a model with a changing level, 
the BIC suggesting a model with a single break corresponding to the year 1898. 
There is a simple explanation for this break: the opening of the Aswan dam in 1898. 
It is worth noting that after modelling the break there is no need for further modelling of 
any dependence about the changing level: the dependence implied by the fitted AR(1) process (left panel), 
with an autoregressive parameter of  $0.51$, is spurious and stems from the neglected data feature of a changing level.

\section{Dating changes in a commodity price series}\label{sec:oilprice}

\begin{figure}[h!]
\begin{center}
\includegraphics{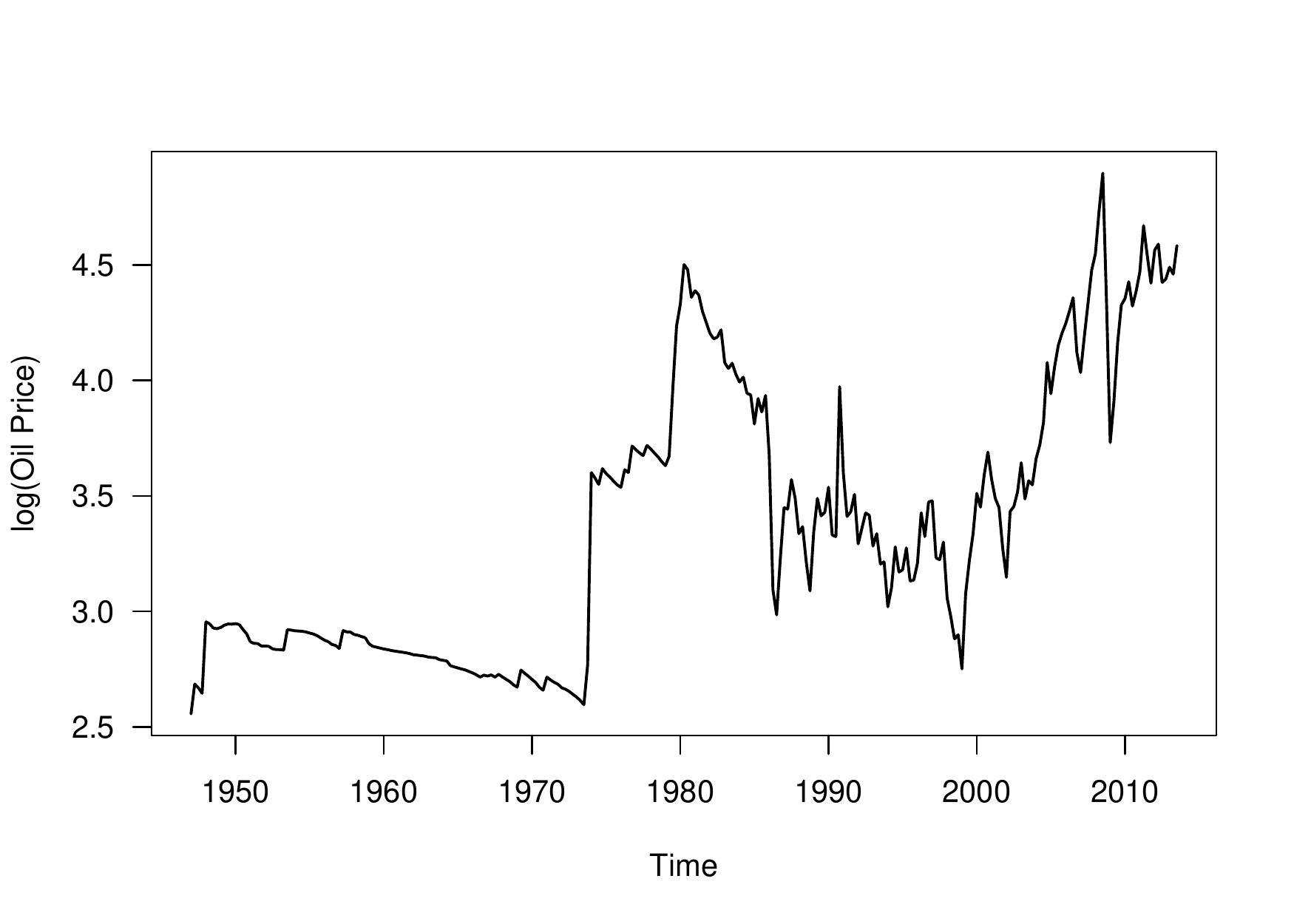}
\caption{\label{fig:wti} Quarterly spot prices for West Texas Intermediate (WTI). Source: Federal Reserve Bank of St. Louis (FRED databse).}
\end{center}
\end{figure}

This section revisits an empirical example presented in 
\cite{compsyn:Zeileis+Kleiber+Kraemer:2003}, namely dating structural breaks in a time series of oil prices. That paper considered a quarterly index of import prices of petroleum products obtained from the
German Federal Statistical Office -- hereafter referred to as the German oil price data -- 
for the period 1960(1) to 1994(4) (base year: 1991). 
The present paper uses a much longer series, a quarterly time series of spot prices for West Texas Intermediate (WTI) -- hereafter referred to as the WTI data -- from 1947(1) to 2013(3).  
It is publicly available from the FRED database of the Federal Reserve Bank of St. Louis, 
more specifically from \url{https://research.stlouisfed.org/fred2/series/OILPRICE/}. 
The series is deflated using the GDP deflator (base year: 2009), which is available from \url{https://research.stlouisfed.org/fred2/series/GDPDEF/}. This deflated version is given in Figure~\ref{fig:wti} 
(data are in logarithms). The task is to compare a change point model for the WTI data with 
the corresponding segmentation for the older German oil price data, 
and also to try out several more recent dating algorithms, on which more below.

\setkeys{Gin}{width=0.8\textwidth}

\begin{figure}[htbp]
\begin{center}
\includegraphics{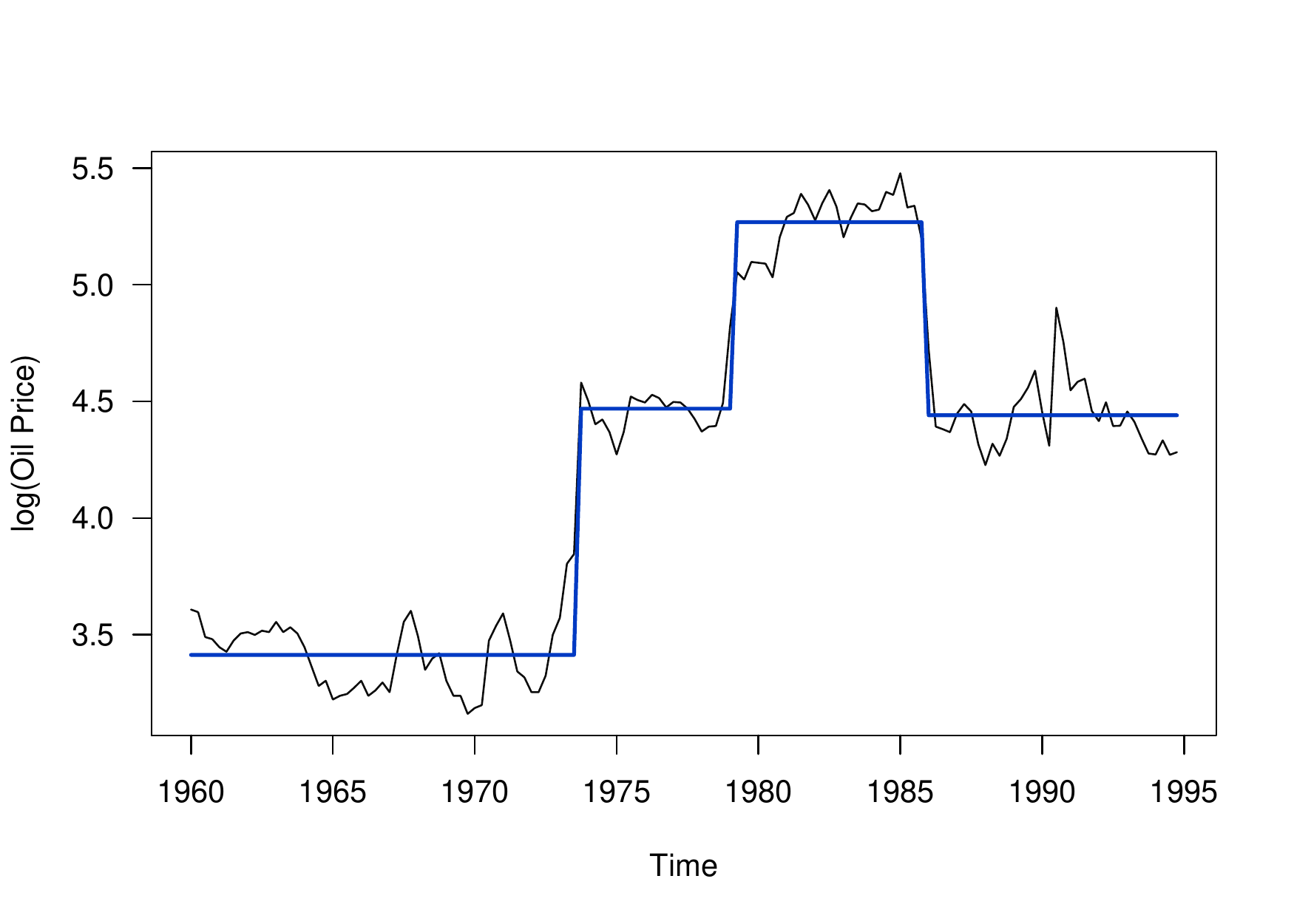}
\caption{\label{fig:goil} German oil price data, 1960(1)
to 1994(4) (base year: 1991). Data source: German Federal Statistical Office. 
Segmentation as in \cite{compsyn:Zeileis+Kleiber+Kraemer:2003}.}
\end{center}
\end{figure}

For ease of reference, the older series along with a segmentation with three regimes 
is given in Figure~\ref{fig:goil} (data are again in logarithms). 
The three breaks are for the quarters 1973(3), 1979(1) and 1985(1). 
The first two breaks correspond to two major historical events, the first oil crisis (the Arab oil embargo 
following the Yom Kippur war) and the beginning of the Iranian revolution.  
The break in 1985(1) may be seen as resulting from demand shifts, 
quarrels within OPEC, and the entry of several new suppliers (namely Great Britain, Mexico, and Norway) 
in international oil markets \citep{compsyn:Zeileis+Kleiber+Kraemer:2003}.


\setkeys{Gin}{width=0.8\textwidth}



\begin{figure}[h!]
\begin{center}
\includegraphics{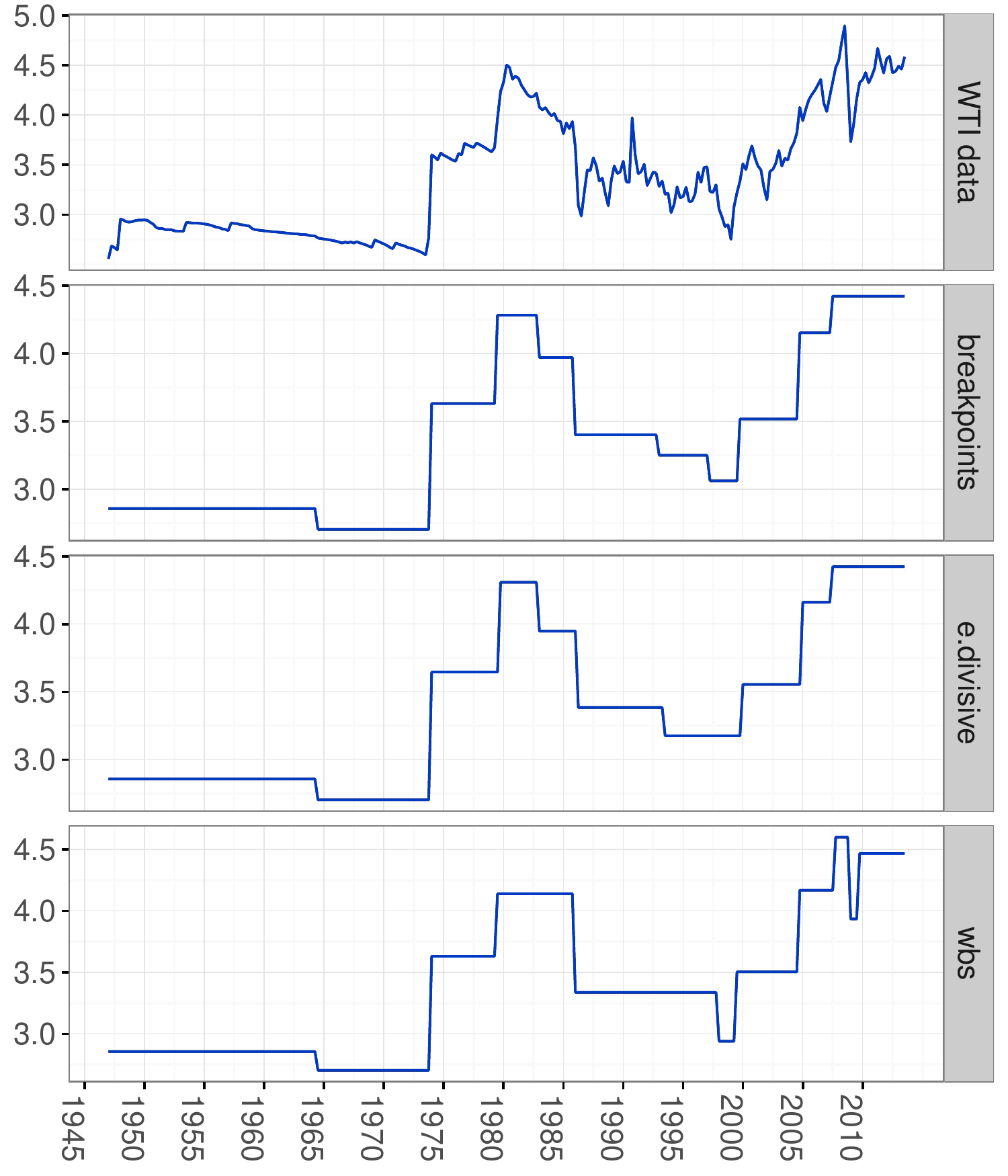}
\caption{\label{fig:wti-compare} WTI data (top panel) and segmentations obtained from three algorithms: 
\texttt{breakpoints}, \texttt{e.divisive} and \texttt{wbs}.}
\end{center}
\end{figure}

Repeating the exercise with the newer WTI data and a minimal segment size of 10 quarters, the BIC now favors a segmentation with 10 regimes. 
The resulting solution is provided in the second panel of Figure~\ref{fig:wti-compare}. 
There is good agreement for the breaks corresponding to major historical events, here estimated at 
1973(4) and 1979(2). Beginning in the second half of the 1990s, there is a gradual trend in the newer series 
that is not described well by a step function. Also, the two oil price series differ visibly 
in the first half of the 1980s, 
leading to two estimated change points for the newer series, at 1982(4) and 1985(4). 
These differences result from the fact that the older series is a price index while the newer is for a single product; 
also, there appear to be exchange rate effects in the older series.

We next compare the least-squares-based solution with two recent methods. 
The first method \citep{compsyn:Matteson+James:2014} uses ideas from cluster analysis 
combined with a nonparametric form of ANOVA based on so-called \emph{energy statistics} 
\citep{compsyn:Rizzo+Szekely:2010}. 
The latter are functions of distances between statistical observations 
in Euclidean spaces (and beyond), the name derives from an analogy with Newton's gravitational potential energy.  
It should be noted that this method assesses differences in entire distributions, 
not just level shifts. However, there is a variant that assesses only changes in the mean; 
the relevant settings for this variant are used below. 
An implementation is available in the \textsf{R} package \pkg{ecp} 
\citep{compsyn:James+Matteson:2014}. The package offers several methods, here only the algorithm 
named \texttt{e.divisive} there is used, a form of hierarchical clustering. 
The second method is \emph{wild binary segmentation} \citep{compsyn:Fryzlewicz:2014} (hereafter: WBS), 
a stochastic algorithm that uses ideas from the wavelets literature. The setup analyzed in the original paper is (\ref{datingmodel}) with i.i.d. Gaussian noise. An implementation is available in the \textsf{R} package \pkg{wbs}. 
 
All three procedures require specification of a trimming parameter, for the least-squares approach 
and \texttt{e.divisive} this is the minimal segment length (or `cluster size', 
in the terminology of the \pkg{ecp} package), details differ from method to method. 
For \texttt{e.divisive}, the minimal segment size was again set to 10 quarters, 
here yielding 9 breaks. For WBS, which does not need a minimal segment size,
the maximum number of breaks was fixed at 10 for comparability reasons. 
The solutions are provided in the third and fourth panel of Figure~\ref{fig:wti-compare}. 
There is good agreement for the major historical events, while the algorithms differ somewhat 
for the second half of the series. This partly reflects the problems with this part of the series 
mentioned above. Overall, WBS tries to adapt to smaller details towards the end of the series. 
It is also worth noting that, using the settings recommended by the authors of the software, 
the WBS algorithm favors a solution with no fewer than 35 
breaks. Clearly, not all of the estimated breaks will be of economic interest. 
This appears to be a problem in some financial applications, where alarms can be frequent with long series. As an example, not all changes identified in \cite{compsyn:Fryzlewicz:2014} for the S\&P 500 index will likely be of practical relevance.

Figure~\ref{fig:wti-compare} provides an overall comparison of all solutions. 
The display highlights similarities and differences of the solutions obtained from the algorithms. 
Notably the \texttt{breakpoints} and \texttt{e.divisive} solutions are very similar. In contrast, 
\texttt{wbs} is more faithful to the more lively part towards the end of the series. 
\texttt{breakpoints} has one further change point in the second half of the 1990s (namely for the quarter 1997(1)), 
the other breaks differ, with one exception, by at most two quarters. 
For example, the break corresponding to the first oil crisis is in 1973(4) according to 
\texttt{breakpoints} and in 1974(1) according to \texttt{e.divisive}. 
For the Iranian revolution break, the algorithms give 1979(2) and 1979(4), respectively.

\setkeys{Gin}{width=0.78\textwidth}

\section{Dating changes in the volatility of a stock market index}
\label{sec:financial}

The previous section considered changes in the mean of a time series. 
It is also possible to study changes in other characteristics of the data, for example variances or autocorrelations. With financial time series, for example stock returns, 
assessing risks is a central issue; both squared and absolute returns may be seen as measures of risk. 
Assessing changes in such transformed returns can be viewed as an indirect check of structural change 
in GARCH-type models of volatility. 

As a brief empirical illustration, we revisit an example from \cite{compsyn:Andreou+Ghysels:2002}. 
They consider four stock market indices (FTSE, Hang Seng, Nikkei, S\&P500) with an eye on changes 
associated with the Asian and Russian financial crises in the second half of the 1990s. 
Here we just consider one of these series, the Hang Seng index for the period 1989--01--04 to 2001--10--19, giving $T = 3338$ observations. For comparability reasons, the data are taken from Datastream. The data for the segmentation algorithm are the Hang Seng absolute returns (for which more breaks are found than for the more common squared returns). 
The original paper documents only the maximal number of breaks used but not the minimal segment size. 
Here we use a minimal segment size corresponding to 10\% of the length of the series, 
which should permit recovery of the segmentation from \cite{compsyn:Andreou+Ghysels:2002}. 
This is only partly possible, however.

\begin{figure}[th!]
\begin{center}
\includegraphics{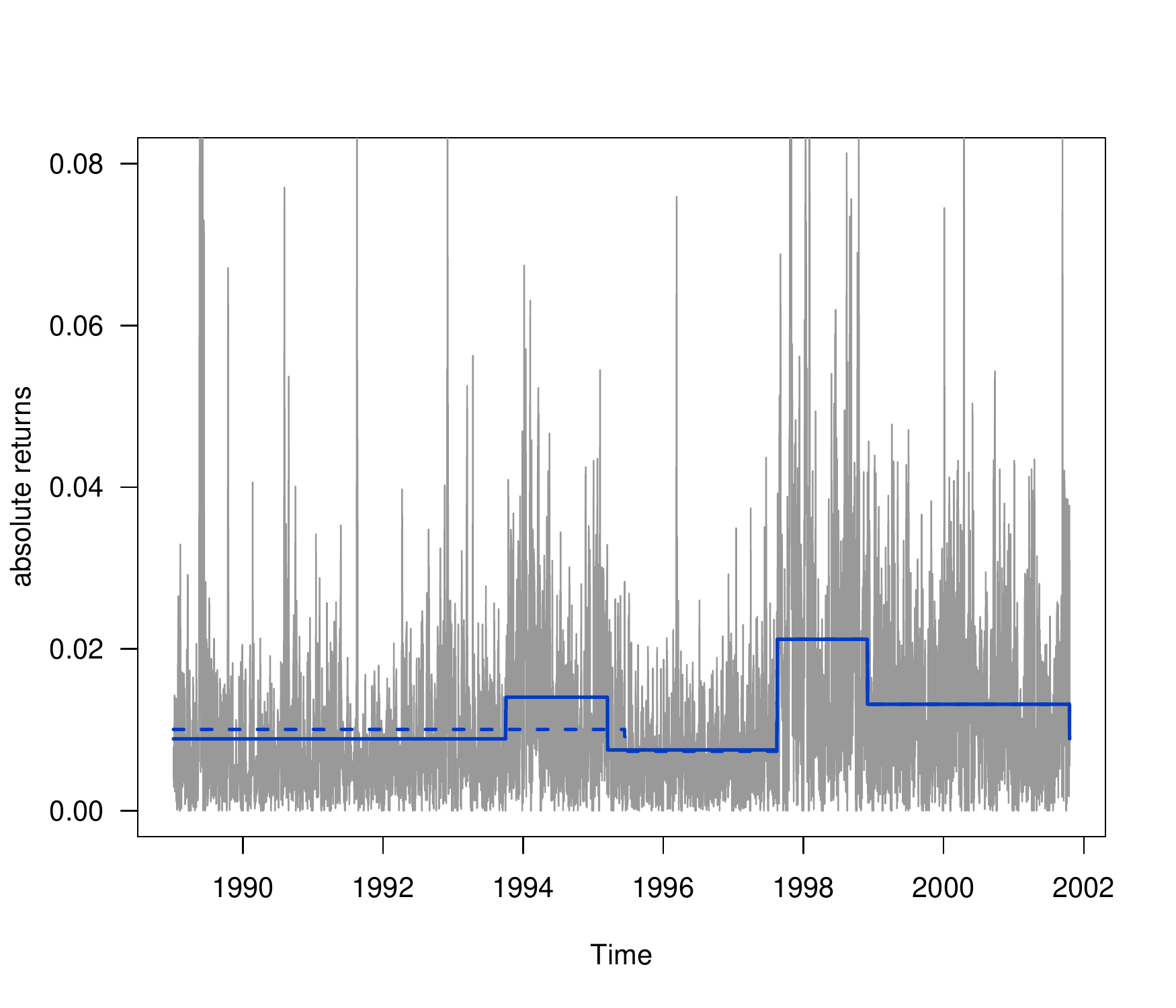}
\caption{\label{fig:vola-plot} Segmentation of Hang Seng absolute returns with \texttt{breakpoints}. 
Dashed line: maximal number of breaks 3, minimal segment size 10\%. 
Solid line: maximal number of breaks 5, minimal segment size 10\%.}
\end{center}
\end{figure}

Figure~\ref{fig:vola-plot} provides a plot of the absolute returns along with two models. The original paper suggests that a segmentation with three breaks, for the dates 1992-07-03, 1995-01-24 and 1997-08-15, 
is optimal. The 3-breaks solution found by \texttt{breakpoints} (the dashed line in the plot) differs, 
it finds 1995-06-14, 1997-08-15 and 1998-11-30, so only the last break from the paper is recovered. Also, the \texttt{breakpoints} solution has no break prior to 1995 but a new break after 1997.

It is also worth noting that the method finds a further break upon increasing the maximum number 
of breaks. Setting the latter to five breaks, a segmentation with four breaks is found. 
Interestingly, the new break at 1993-10-01 is before 1995, although not overly close to 
the 1992-07-03 break of \cite{compsyn:Andreou+Ghysels:2002}. 
Further experiments with the minimal segment length (down to 5\% of the length of the series) 
and the admissible number of breaks (up to 10) suggest that the results are quite sensitive to 
the settings of these parameters. The only break that is practically always found is for August 1997, 
it is often estimated at 1997-08-15 and is associated with the Asian financial crisis.

\setkeys{Gin}{width=0.8\textwidth}

\section{Discussion and outlook}

This chapter has illustrated some basic ideas in change point analysis. The exposition was confined to retrospective methods for univariate time series. (Most of) The methods described have extensions to regression models \citep{compsyn:Perron:2006}. Many further topics had to be excluded, notably the timely topic of on-line monitoring and also structural change in multivariate or functional data. For on-line monitoring, in some fields referred to as surveillance or `quickest detection' problems, see the survey by \cite{compsyn:Frisen:2009} and references therein, for an exposition of associated optimality issues 
in a financial setting see \cite{compsyn:Shiryaev:2002}. Multivariate and functional data are briefly addressed 
in the recent survey by \cite{compsyn:Horvath+Rice:2014}, where further references may be found.

The literature will likely continue to grow rapidly, for several reasons: 
The classical methods are largely confined to linear models fitted via least squares methods. 
Nonlinear models for discrete-valued data are needed in some applications, 
but here the literature is still relatively small. Also, the growing number of large data sets 
demands improvements on the algorithmic side. Unfortunately, many recent methods are not readily 
available in statistical software, which to some extent hinders progress. 
A further big challenge is the unification of this widely scattered literature.

\section*{Computational Details} \label{sec:computational}

All results were obtained using
\proglang{R}~3.2.4, 
with the packages
\pkg{strucchange}~1.5-1,
\pkg{ecp}~2.0.0,
and
\pkg{wbs}~1.3,
on PCs running Mac OS X, version~10.10.5. 
Some plots were drawn using the package \pkg{ggplot2}~2.1.0 
\citep{compsyn:Wickham:2009}.

\bibliography{compsyn}

\end{document}